\documentclass[12pt]{iopart}
\usepackage{amssymb}
\usepackage{graphics}
\usepackage{graphicx}

\begin{document}

\title{Noncommutative chaotic inflation and WMAP three year results}
\author{
Xin Zhang${}^{a,b,c}$ and Feng-Quan Wu${}^{a,b}$}

\address{${}^{a}$ CCAST (World Laboratory),
P.O.Box 8730, Beijing 100080, People's Republic of China}
\address{${}^{b}$ Institute of High Energy Physics, Chinese
Academy of Sciences, P.O.Box 918(4), Beijing 100049, People's
Republic of China}
\address{${}^{c}$ Interdisciplinary Center of Theoretical Studies, Chinese
Academy of Sciences, P.O.Box 2735, Beijing 100080, People's Republic
of China}
\date{\today}

\begin{abstract}

Noncommutative inflation is based upon the consideration of some
effects of the space-time uncertainty principle motivated by ideas
from string/M theory. The CMB anisotropies may carry a signature of
this very early Universe correction from the space-time uncertainty
principle and can be used to place constraints on the parameters of
the noncommutative inflation model. In this paper we analyze the
noncommutative chaotic inflation model by means of the WMAP three
year results. We show that the noncommutative chaotic inflation
model can produce a large negative running of spectral index within
a reasonable range of $e$-folding number $N$, provided that the
value of $p$ in the potential $V(\phi)\propto \phi^p$ is enhanced
roughly to $p\sim 12-18$. Applying a likelihood analysis to the WMAP
results, we find the best-fit values of parameters for the
noncommutative chaotic inflation model. In addition, this model
predicts a rather big gravitational wave which may be tested by the
upcoming experiments.

\end{abstract}

\bigskip

Einstein's general relativity will break down at very high energies
in the early Universe when quantum effects are expected to be
important. If inflation \cite{Guth} happens at the very early time
and if the period of inflation lasts sufficiently long, then the
effects of quantum gravity should in principle leave an imprint on
the primordial spectrum of perturbations, since the wavelengths of
perturbations emerged from short distances in the early stages of
inflation are stretched to the cosmic scales observable today by the
rapid expansion during inflation. Though we lack a complete theory
of quantum gravity presently, we can still look for the hint of
these quantum gravity imprints through cosmological experiments such
as WMAP and SDSS. Since string/M theory is a promising framework for
quantum gravity, it is of interest to explore specific stringy
corrections to the spectrum of fluctuations. Many authors have
considered and discussed this problem. For the first discussion on
the trace of the trans-planckian effect in the primordial spectrum
of perturbations see \cite{initial}; for a review containing a
comprehensive list of references see \cite{review}. In particular,
it should be pointed out that the possible effects of quantum
gravity in the spectra of fluctuations in inflationary cosmology may
be induced by the consequences of a basic stringy effect, namely the
space-time uncertainty relation \cite{uncpr}
\begin{equation} \label{stunc}
\Delta t \Delta x_{\rm phys}\geqslant L_s^2,
\end{equation}
where $t$, $x_{\rm phys}$ are the physical space-time coordinates
and $L_s$ is the string scale. It was shown in
\cite{Alexander:2001dr} that this space-time uncertainty principle
may yield inflation from pure radiation. A more modest approach was
pioneered in \cite{Brandenberger:2002nq}, where the consequences
were studied by imposing Eq.(\ref{stunc}) into the action for
cosmological perturbations on an inflationary background. On a
general ground, if inflation is indeed affected by physics at a
scale close to string scale or a related scale, it can be expected
that space-time uncertainty must leave traces in the CMB power
spectrum \cite{Brandenberger:2002nq,Chu:2000ww,Lizzi:2002ib}. The
primordial power spectrum based upon the noncommutative inflation
model \cite{Brandenberger:2002nq} has been given in
\cite{Huang:2003zp,Tsujikawa:2003gh,Huang:2003hw,Huang:2003fw}, and
the explicit calculation indeed shows that the effects can be
observed.

The standard concordance $\Lambda$CDM model, which can arise from an
inflationary background cosmology in which the quasi-exponential
expansion of space is driven by a scalar field, still provides a
fairly good fit to the recent three-year WMAP \cite{WMAP3yr} and
earlier observations. Assuming that the primordial fluctuations are
adiabatic with a power law spectrum, the WMAP three-year data
require a spectral index that is significantly less than the
Harrison-Zel'dovich-Peebles scale-invariant spectrum ($n_s=1$,
$r=0$). This suggests, for power law inflationary models, a
detectable level of gravity waves. However, many inflationary models
can only predict a much smaller gravity wave amplitude. Moreover,
the WMAP three-year data can place significant constraints on
inflationary models. For example, the chaotic inflationary model
with the inflaton potential, $V(\phi)\propto \phi^p$, has been
tested in light of the current data, with the result that the data
prefer the $m^2\phi^2$ model over the $\lambda\phi^4$ model,
assuming a power-law primordial power spectrum. When allowing for a
running spectral index, the data will favor a large negative running
index and a large tensor amplitude (characterized by $r$, the ratio
of the tensor to scalar power spectrum). However, it is rather hard
to produce a large absolute value of the running spectral index
within the framework of the usual slow-roll inflationary models
\cite{hardrun}. Hence, the confirmation of this suggestive trend is
very important for our understanding of the early Universe physics,
especially of the quantum gravity relevant physics.

It has been mentioned that the CMB anisotropies may carry a
signature of the very early Universe corrections from the space-time
uncertainty principle motivated by string theory, and can be used to
place constraints on the parameters appearing in the noncommutative
inflation model. A recent paper \cite{Huang:2006um} shows that the
noncommutative inflation model can nicely produce a large running
spectral index. In \cite{Huang:2006um}, some concrete noncommutative
inflation models such as the chaotic inflation and power-law
inflation have also been re-examined. It is well known that the
usual chaotic inflation model can only predict the negligible amount
of running index, $dn_s/d\ln k\thickapprox -10^{-3}$. Though the
noncommutative chaotic inflation model can realize a considerable
running index, for the $m^2\phi^2$ model and the $\lambda\phi^4$
model, the authors of \cite{Huang:2006um} find that a low number of
$e$-folds, say $N\sim 10$, is required. Needless to say, this is
fairly unnatural since such low values of $N$ may not be sufficient
to resolve the naturalness problems such as the flatness and the
horizon puzzles. A reasonable range of $N$ has been derived in
\cite{Alabidi:2005qi}, that is $47<N<61$ (or $N=54\pm 7$). In this
paper, we will show that the noncommutative chaotic inflation model
can provide a large negative running spectral index within the
reasonable range of $N$, provided that the value of $p$ in the
potential $V(\phi)\propto \phi^p$ is enhanced, say $p\sim 15$. In
addition, we perform a likelihood analysis to find the best-fit
values of the parameters of the noncommutative chaotic inflation
model to the WMAP results of the spectral index and its running.
Thus we are able to constrain the noncommutative model so as to
place limits on the space-time noncommutativity in the string scale.

The space-time noncommutative effects can be encoded in a new
product, star product, replacing the usual algebra product. The
evolution of a homogeneous and isotropic background will not change
and the standard cosmological equations based upon the
Friedmann-Robertson-Walker (FRW) metric remain the same:
\begin{equation}
\ddot{\phi}+3H\dot{\phi}+V'(\phi)=0,
\end{equation}
\begin{equation}
H^2=\left(\dot{a}\over a\right)^2={1\over 3M_{\rm
Pl}^2}\left({1\over2}\dot{\phi}^2+V(\phi)\right),
\end{equation}
where $M_{\rm Pl}$ is the reduced Planck mass. If $\dot{\phi}^2\ll
V(\phi)$ and $\ddot{\phi}\ll 3H\dot{\phi}$, the scalar field shall
slowly roll down its potential. Define some slow-roll parameters,
\begin{equation}
\epsilon=-{\dot{H}\over H^2}={M_{\rm Pl}^2\over 2}\left(V'\over
V\right)^2,
\end{equation}
\begin{equation}
\eta=\epsilon-{\ddot{H}\over 2H\dot{H}}=M_{\rm Pl}^2{V''\over V},
\end{equation}
\begin{equation}
\xi^2=7\epsilon\eta-5\epsilon^2-2\eta^2+\zeta^2=M_{\rm
Pl}^4{V'V'''\over V^2},
\end{equation}
where $\zeta^2=\stackrel{...}{H}/(2H^2\dot{H})$, then the slow-roll
condition can be expressed as $\epsilon, \eta\ll 1$.

The stringy space-time uncertainty relation leads to changes in the
action for the metric fluctuations. The action for scalar metric
fluctuations can be reduced to the action of a real scalar field $u$
with a specific time-dependent mass which depends on the background
cosmology. Thus the modified action of the perturbation led by the
stringy space-time uncertainty relation can be expressed as
\cite{Brandenberger:2002nq}
\begin{equation}
S=V_T\int\limits_{k<k_0}d\tilde{\eta}d^3k{1\over
2}z_k^2(\tilde{\eta})(u'_{-k}u'_k-k^2 u_{-k}u_k),
\end{equation}
where $V_T$ is the total spatial coordinate volume, a prime denotes
the derivative with respect to the modified conformal time
$\tilde{\eta}$, $k$ is the comoving wave number, and
\begin{equation}
z_k^2(\tilde{\eta})=z^2(\beta_k^+\beta_k^-)^{1/2},~~~~z={a\dot{\phi}\over
H},
\end{equation}
\begin{equation}
{d\tilde{\eta}\over
d\tau}=\left(\beta_k^-\over\beta_k^+\right)^{1/2},~~~~\beta_k^\pm={1\over
2}(a^{\pm 2}(\tau+kL_s^2)+a^{\pm 2}(\tau-kL_s^2)),
\end{equation}
here $L_s$ is the string length scale,
$k_0=(\beta_k^+/\beta_k^-)^{1/4}L_s$ and $\tau$ denotes a new time
variable in terms of which the stringy uncertainty principle takes
the simple form $\Delta\tau\Delta x\geqslant L_s^2$, using comoving
coordinates $x$. The case of general relativity corresponds to
$L_s=0$.

In \cite{Huang:2003fw}, the power spectrum in the noncommutative
space-time of the curvature perturbation, ${\cal R}\equiv
\Phi+H\delta\phi/\dot{\phi}$, where $\Phi$ is the scalar metric
perturbation in longitudinal gauge, has been calculated by
evaluating the time when fluctuations are generated,
\begin{equation}
{\cal P_R}\simeq {1\over 2\epsilon}{1\over M_{\rm Pl}^2}\left(H\over
2\pi\right)^2(1+\mu)^{-4-6\epsilon+2\eta}={V/M_{\rm Pl}^4\over
24\pi^2\epsilon}(1+\mu)^{-4-6\epsilon+2\eta},
\end{equation}
where $\mu=H^2k^2/(a^2M_s^4)$ is the noncommutative parameter, $H$
and $V$ takes the values when the fluctuation mode $k$ crosses the
Hubble radius, and $M_s=L_s^{-1}$ is the string mass scale. Then the
spectral index of the power spectrum of the scalar fluctuations and
its running can be given,
\begin{equation}\label{index}
n_s-1\equiv s={d\ln {\cal P_R}\over d\ln
k}=-6\epsilon+2\eta+16\epsilon\mu,
\end{equation}
\begin{equation}\label{running}
{d n_s\over d\ln k}\equiv
\alpha_s=-24\epsilon^2+16\epsilon\eta-2\xi^2-32\epsilon\eta\mu.
\end{equation}
When $L_s\rightarrow 0$ or $M_s\rightarrow +\infty$, the
noncommutative parameter $\mu=H^2k^2/(a^2M_s^4)\rightarrow 0$, Eqs.
(\ref{index}) and (\ref{running}) reproduce the results of general
relativity in the commutative case. The tensor-scalar ratio is also
given,
\begin{equation}\label{tensor}
r=16\epsilon.
\end{equation}
On the other hand, the best fit results of the three-year WMAP data
only for the $\Lambda$CDM model with both a running spectral index
and tensor modes are \cite{WMAP3yr}
\begin{equation}\label{scalarWMAP}
n_s=1.21^{+0.13}_{-0.16},~~~~dn_s/d\ln k=-0.102^{+0.050}_{-0.043},
\end{equation}
and
\begin{equation}\label{tensorWMAP}
r\leq 1.5~~~{\rm at~95\%~CL}.
\end{equation}
The fit of the WMAP data  shows that a large negative running
spectral index is favored by the data, and if allowing for a running
index, then models with large tensor components are consistent with
the data.

The chaotic inflation has the potential of the form $V(\phi)=\lambda
M_{\rm Pl}^{4-p}\phi^p$ with $p$ a positive integer and $\lambda$
dimensionless, which provides a way of descending from the Planck
scale with chaotic initial field conditions \cite{chaotic}. The
slow-roll parameters are related to the number of $e$-folds $N$ of
inflation between the epoch when the horizon scale modes left the
horizon and the end of inflation by
\begin{equation}
\epsilon={p\over 4N},~~~~ \eta={p-1\over 2N},
~~~~\xi^2={(p-1)(p-2)\over 4N^2}.
\end{equation}
In the noncommutative case, the spectral index and its running as
well as the tensor-scalar ratio can be expressed as
\begin{equation}
n_s=1+\left(\mu-{1\over 8}-{1\over 4p}\right)r,
\end{equation}
\begin{equation}
\alpha_s=-{p+2\over 32p^2}r^2\left(1+{8p(p-1)\over p+2}\mu\right),
\end{equation}
\begin{equation}
r={4p\over N}.
\end{equation}
It can be seen that the running of the spectral index is always
negative for the chaotic inflation model. Moreover, the power
spectrum is always a red spectrum in the usual commutative case. But
it will become blue if $\mu > (p+2)/8p$ in the noncommutative
chaotic inflation. For the chaotic inflation in the usual
commutative case, the amount of the running of the spectral index is
negligible, $\alpha_s\sim -10^{-3}$. Thus in \cite{WMAP3yr}, this
model in the quadratic case $m^2\phi^2$ and in the quartic case
$\lambda\phi^4$ have been tested by using three-year WMAP data
assuming a power-law primordial power spectrum, with the result that
the data rule out rather firmly $p\geq 4$. However, the chaotic
inflation in the noncommutative case will be capable of producing a
large enough running of the spectral index provided that choosing a
noncommutative parameter $\mu$ properly, as shown in
\cite{Huang:2006um}. The authors of \cite{Huang:2006um} only
analyzed the simplest cases of noncommutative chaotic inflation
model, i.e. the cases of $p=2$ and $p=4$, and found that in order to
realize a large running of the index for matching the WMAP data
roughly, a rather low $e$-folding number $N$ is required, say $N\sim
10$. Such a low value of $N$, however, seems fairly unnatural as
confessed in \cite{Huang:2006um}. Actually, reasonable value of $N$
should be around 50, as demonstrated in \cite{Alabidi:2005qi},
$N=54\pm 7$. We will show that in order to reconcile this conflict
within the noncommutative framework the model parameter $p$ should
be enhanced to $p\sim 12-18$.

\begin{figure}[htbp]
\begin{center}
\includegraphics[scale=1.1]{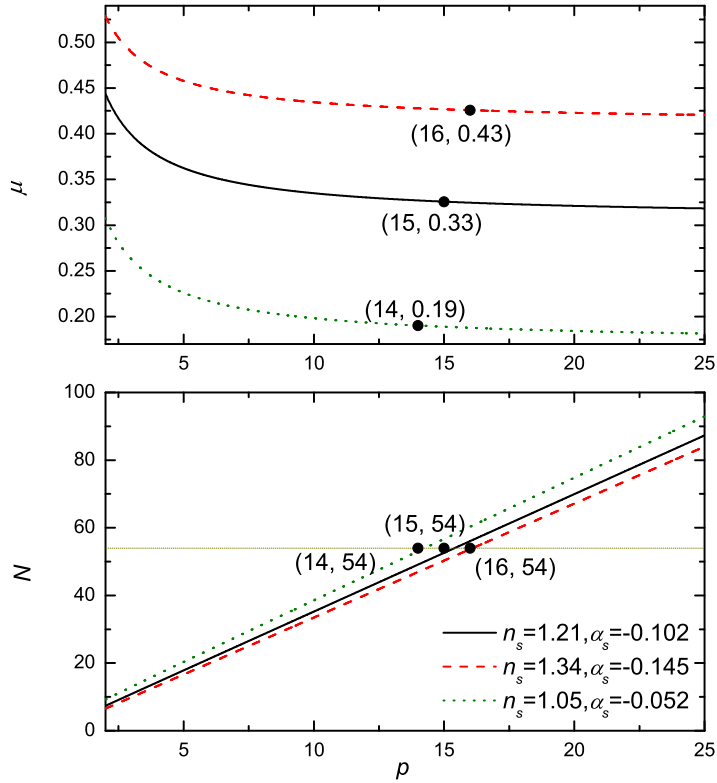}
\caption[]{\small The constraint relationship curves for $\mu(p)$
and $N(p)$. Points locate the coordinates corresponding to $N=54$.
Note that we restrict $p$ to be an integer.}\label{fig:constraints}
\end{center}
\end{figure}

\begin{figure}[htbp]
\begin{center}
\includegraphics[scale=1.05]{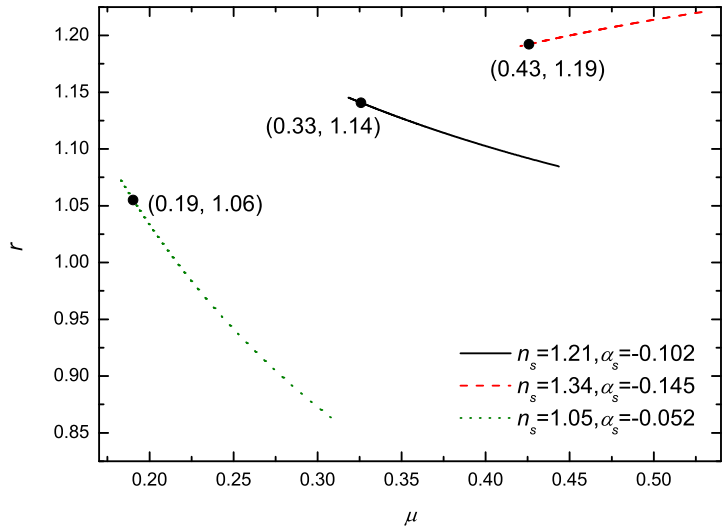}
\caption[]{\small The relationship between $\mu$ and $r$. Points
locate the coordinates corresponding to $N=54$.}\label{fig:rmu}
\end{center}
\end{figure}

\begin{figure}[htbp]
\begin{center}
\includegraphics[scale=1.2]{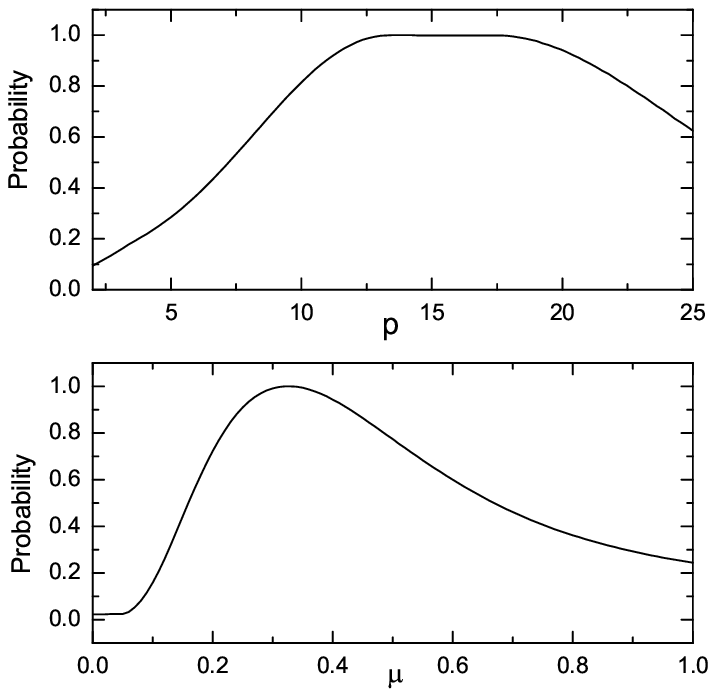}
\caption[]{\small Probability distributions of the likelihood
analysis for the parameters $p$ and $\mu$, assuming that $47<N<61$.
The best fit happens at $p=13.5$, $\mu=0.33$ and
$N=47.56$.}\label{fig:likelihood}
\end{center}
\end{figure}

Utilizing the noncommutative chaotic inflation results (\ref{index})
and (\ref{running}) to fit the fixed values of spectral index and
its running, for instance the WMAP central values $n_s=1.21$ and
$\alpha_s=-0.102$, gives a constraint relationship between the
noncommutative parameter $\mu$ and the potential parameter $p$,
namely $\mu(p)$, and then obtains the constraint relationship of
$N(p)$. This case is plotted in figure \ref{fig:constraints}, as
indicated by the black solid lines. The plot explicitly shows that
the value of $N$ increases with the value of $p$, and $N(p)$ is
nearly linear. We also see that $\mu(p)$ becomes nearly flat when
$p$ goes beyond 10. The relation between the tensor-scalar ratio $r$
and the noncommutative parameter $\mu$ in this case can also be
obtained, as shown in figure \ref{fig:rmu}, the black solid line. We
see that the noncommutative chaotic inflation model can predict a
rather big gravitational wave, $r\sim 1$, which is within the range
of the WMAP and can be confirmed by the upcoming experiments
eventually. If  the $e$-folds $N$ is taken to be a reasonable value,
say 54, then we get $p=15$, $\mu=0.33$ and $r=1.14$, as shown in
figures \ref{fig:constraints} and \ref{fig:rmu}. Note that here $p$
is restricted to be an integer and is not needed to be an even
number. Furthermore, we also consider the uncertainties of $n_s$ and
$\alpha_s$ in (\ref{scalarWMAP}) simply. We take two cases into
account, $n_s=1.34$ and $\alpha_s=-0.145$ as well as $n_s=1.05$ and
$\alpha_s=-0.052$, and show the results in figures
\ref{fig:constraints} and \ref{fig:rmu}, represented by the red
dashed lines and green dotted lines, respectively. According to
$N=54$, we have, for these two cases respectively, $p=16$,
$\mu=0.43$ and $r=1.19$ as well as $p=14$, $\mu=0.19$ and $r=1.06$.
Hence, fitting the WMAP three year results and at the same time
considering a reasonable $e$-folds $N$, we conclude that the
noncommutative chaotic inflation model should have a potential
parameter $p\sim 14-16$. What is more, the noncommutative chaotic
inflation predicts a rather big gravitational wave, $r\sim 1$.

Finally, we perform a likelihood analysis to find the best-fit
values of the model parameters, including $p$, $\mu$ and $N$, to the
WMAP three-year results. According to the WMAP results of the
spectral index and its running (\ref{scalarWMAP}), the likelihood
values of model parameters can be derived by evaluating the
$\chi^2$-distribution. First, we assume a rather large range for the
$e$-folds $N$, $14<N<75$. The likelihood analysis shows that the
best fit happens at $p=20.86$, $\mu=0.32$ and $N=72.87$. Then, we
assume a more reasonable range for the $e$-folds, $47<N<61$. The
best fit for this case happens at $p=13.5$, $\mu=0.33$ and
$N=47.56$. The probability distributions of the likelihood values of
$p$ and $\mu$ are shown in figure \ref{fig:likelihood}. In a word,
the WMAP three year results require a large $p$ for the
noncommutative chaotic inflation.

In summary, we analyze the noncommutative chaotic inflation model by
means of the WMAP three year results. Noncommutative inflation is a
modification of standard general relativity inflation which takes
into account some effects of the space-time uncertainty principle
motivated by ideas from string theory. This effect of quantum
gravity of the very early Universe may in principle lead to some
corrections to the primordial power spectrum of perturbations and
thus leave an imprint in the CMB anisotropy measurements. Indeed,
the noncommutative inflation model can account for the WMAP three
year results gracefully. In this paper, we show that the
noncommutative chaotic inflation model can provide a large negative
running of the spectral index within a reasonable range of
$e$-folding number $N$, provided that the value of $p$ in the
potential $V(\phi)\propto \phi^p$ is enhanced roughly to $p\sim
12-18$. In addition, we find that the noncommutative chaotic
inflation model predicts a rather big gravitational wave which may
be tested by the upcoming experiments.


\section*{Acknowledgements}

We are grateful to Qing-Guo Huang and Miao Li for useful
discussions. This work was supported by the Natural Science
Foundation of China (Grant No. 111105035001).


\end{document}